\documentclass{IEEEtran}

\usepackage[left=0.68in,right=0.673in,top=0.75in,bottom=1.1in]{geometry}
\usepackage{lipsum}
\usepackage{amsmath}
\usepackage{graphicx}
\usepackage{color,soul}
\usepackage[export]{adjustbox}
\usepackage{tikz}
\usepackage{pgfplots}
\usepackage{amssymb}
\usepackage[caption=false]{subfig}
\usepackage{caption}
\usepackage[utf8]{inputenc}
\usetikzlibrary{decorations.pathreplacing}
\graphicspath{{images/}}
\usetikzlibrary{patterns}
\usepgfplotslibrary{fillbetween}
\pgfplotsset{compat=1.15,width=10cm}
\usepackage{cite}
\usepackage{optidef}
\usepackage{amsthm}
\usepackage[ruled,linesnumbered]{algorithm2e}
\usepackage{setspace}

\usepackage{booktabs}
\usepackage{multirow}
\usepackage{xcolor}
\usepackage{filecontents}
\usepackage[switch]{lineno}

\usepackage{color,soul}

\title{Integrating Terrestrial and Non-Terrestrial Networks for Sustainable 6G Operations: A Latency-Aware Multi-Tier Cell-Switching Approach}
\author{\IEEEauthorblockN{ Metin Ozturk$^{1}$, Maryam Salamatmoghadasi$^2$, Halim Yanikomeroglu$^2$}\\
\IEEEauthorblockA{ $^1$Electrical and Electronics Engineering, Ankara Yıldırım Beyazıt University, Ankara, Türkiye}\\
\IEEEauthorblockA{$^2$Non-Terrestrial Networks (NTN) Lab, Systems and Computer Engineering, Carleton University, Ottawa, Canada}
}

\begin{document}
\maketitle

\begin{abstract}
Sustainability is paramount in modern cellular networks,  which face significant energy consumption challenges from rising mobile traffic and advancements in wireless technology. Cell-switching, well-established in literature as an effective solution, encounters limitations such as inadequate capacity and limited coverage when implemented through terrestrial networks (TN). This study enhances cell-switching by integrating non-terrestrial networks (NTN)—including satellites (used for cell-switching for the first time), high altitude platform stations (HAPS), and uncrewed aerial vehicles (UAVs)—into TN. This integration significantly boosts energy savings by expanding capacity, enhancing coverage, and increasing operational flexibility. 
We introduce a multi-tier cell-switching approach that dynamically offloads users across network layers to manage energy effectively and minimize delays, accommodating diverse user demands with a context-aware strategy. 
Additionally, we explore the role of artificial intelligence (AI), particularly generative AI (GenAI), in optimizing network efficiency through data compression, handover optimization between different network layers, and enhancing device compatibility, further improving the adaptability and energy efficiency of cell-switching operations.
A case study confirms substantial improvements in network power consumption and user satisfaction, demonstrating the potential of our approach for future networks.


\end{abstract}

\section{Introduction}\label{sec:intro}
Rapid growth in mobile data traffic and the increasing demand for ubiquitous connectivity pose significant challenges to energy consumption of wireless networks. As networks expand to accommodate diverse applications, energy consumption has become a critical concern, particularly with the proliferation of base stations~(BSs) required to support the growing demand. Since BSs account for 60\% to 80\% of a network's total energy usage~\cite{bs_power}, efficient energy management strategies are crucial for achieving sustainable operations.
Cell-switching emerges as a pivotal strategy to address these challenges by dynamically deactivating underutilized or idle BSs, thereby reducing unnecessary energy consumption and operational costs while ensuring network performance. By intelligently redistributing traffic to active BSs, cell-switching helps balance network load, extend the life of the equipment, and minimize operational costs.
Extensive research has demonstrated the potential of cell-switching in terrestrial networks (TN)~\cite{bs_power}, offering promising energy efficiency gains. However, traditional TN-based approaches are inherently limited by coverage constraints and capacity demands, necessitating the integration of additional network layers to enhance flexibility and sustainability.

Integrating non-terrestrial networks (NTN) with TN significantly boosts cell-switching efficiency by offering additional capacity and coverage, reducing the reliance on high-power terrestrial BSs, and enabling more flexible and adaptive resource management. NTN elements such as satellites, high altitude platform stations (HAPS), and uncrewed aerial vehicles (UAVs) can offload traffic from energy-intensive terrestrial BSs during low-traffic periods or in remote areas, thus reducing the overall energy consumption of the network. Unlike TN, NTN platforms—particularly HAPS, and satellites—often leverage renewable energy sources such as solar power, making them more sustainable and reducing the dependence on grid power. Furthermore, NTN facilitates dynamic load balancing by providing coverage in underserved areas, preventing overprovisioning of TN resources, and allowing more efficient utilization of available energy across the multi-tier network.
This integration not only bridges
the digital divide by connecting underserved areas~\cite{itu_vision_november} and
achieving ubiquitous connectivity but also supports rapid
deployment in emergencies and reduces reliance on TN. This holistic approach advances global connectivity and sustainability, aligning with the International Telecommunication Union's (ITU) objectives for future networks~\cite{itu_vision_november}.

UAVs, as the initial layer in NTNs, offer rapid deployment and flexible coverage, adapting to dynamic demands and emergencies, and optimizing network performance where terrestrial support is limited. 
HAPS, positioned at about 20~km altitude, provide persistent connectivity. Recognized by ITU as high-altitude IMT BSs (HIBS)~\cite{itu_vision_november}, its potential is augmented by new frequency bands from the World Radio Communication Conference 2023 (WRC-23)~\cite{WRC_23}, enhancing broadband access with minimal infrastructure. Cited by the World Economic Forum as a top emerging technology~\cite{WEF}, HAPS plays a crucial role in shaping the sixth generation of mobile communications~(6G) and beyond, serving remote areas and aiding disaster recovery efficiently. Satellites, at the highest NTN layer, continually evolve to boost terrestrial and stratospheric communications, with advancements enabling direct-to-device communications that greatly enhance mobile broadband and ensure ubiquitous connectivity.

On the other hand, the integration of NTN introduces complexity in managing latency variations across different elements such as UAV, HAPS, and satellite, making effective management crucial for seamless operation across network layers. Understanding latency dynamics is vital for ensuring that cell-switching conserves energy while maintaining communication reliability and timeliness. 
Additionally, the integration of artificial intelligence (AI), especially generative AI (GenAI), enhances adaptability and efficiency in resource management by converting data into more manageable formats, reducing network load, and improving traffic offloading and cell-switching.

In this article, we introduce a novel, comprehensive multi-tier cell-switching strategy that significantly enhances overall network efficiency by integrating TN and NTN. This innovative approach, being the first to use satellites for cell-switching, incorporates UAVs, HAPS, and satellites, adapting cell-switching to diverse network needs. Each component contributes uniquely: UAVs provide flexibility and rapid response, especially for delay-sensitive applications; HAPS ensures widespread, reliable connectivity without frequent maintenance; and satellites deliver robust global coverage. We present two specific strategies: an energy-focused approach that minimizes power consumption, and a delay-focused approach that reduces latency to improve user satisfaction. 

\begin{figure*}[h!]
    \centering
    \includegraphics[width=.9\textwidth]{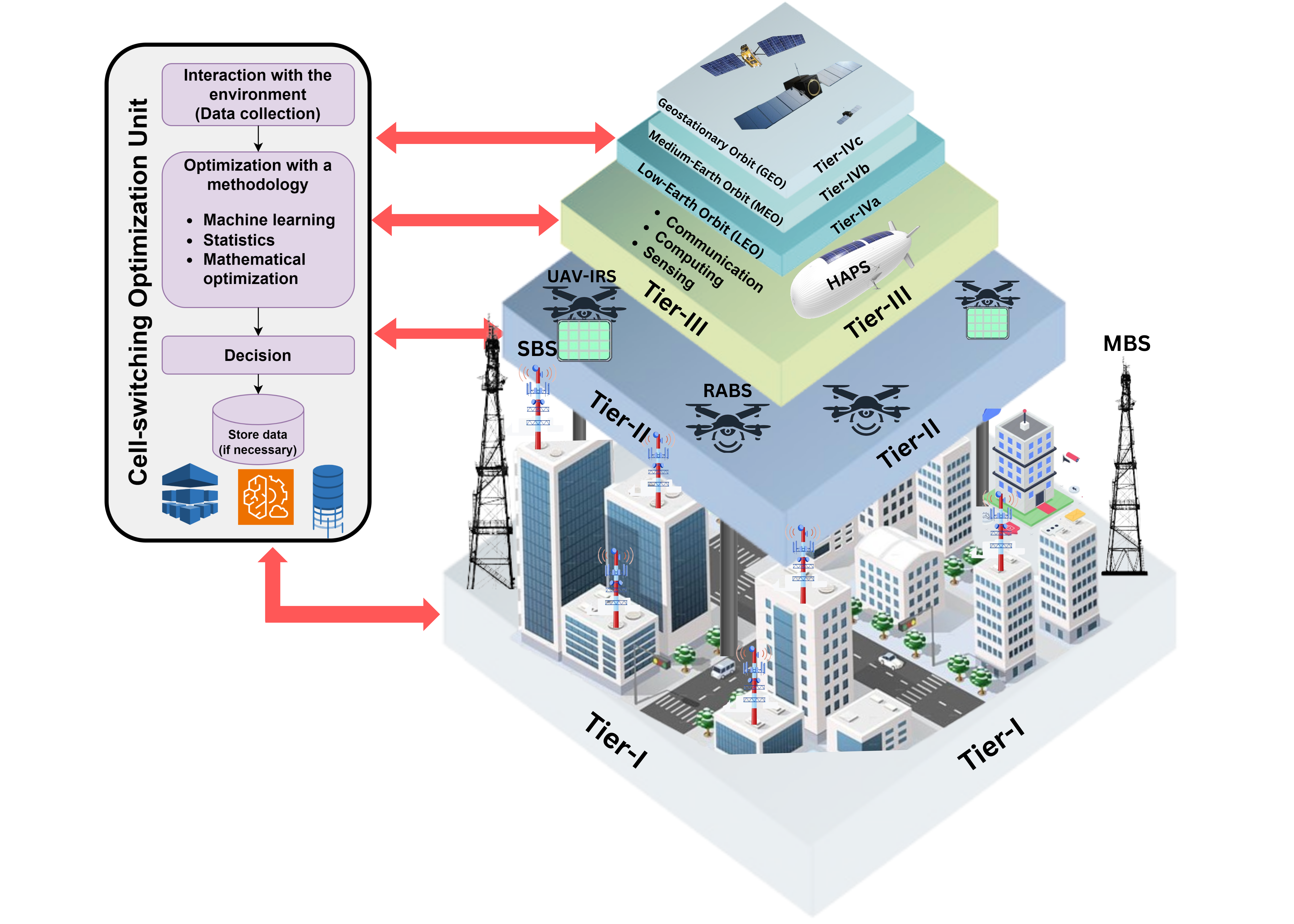}
    \caption{A comprehensive multi-tier cell-switching approach. Tier-I is TN, Tier-II is UAV-based aerial network, Tier-III includes HAPS-SMBS, and Tier-IV is the satellite network.}
    \label{fig:concept}
\end{figure*}
\section{Comprehensive Multi-Tier Cell-Switching}\label{sec:ffmd}
As network demand grows, reflected by an increase in BSs, the potential for energy savings through cell-switching rises. However, this also boosts capacity demands, which can limit the feasibility of traditional cell-switching operations based solely on TN technologies.

To address capacity constraints and enhance cell-switching, integrating multiple network tiers offers several advantages:
\begin{itemize}
    \item \textbf{Capacity and Coverage Expansion:} Multiple tiers handle increased user traffic and offloading demands, spread the load across networks, extend coverage to remote areas, and enhance flexibility in network operations.
    \item \textbf{Deployment Cost Reduction:} Utilizing existing infrastructure like satellites and HAPS for cell-switching avoids high costs, as these elements are already operational.
    \item \textbf{Service Quality Enhancement:} Specializing tiers for different services ensures that latency-sensitive applications receive priority, improving user experience.
\end{itemize}

\subsection{Capacity is Everything in Cell-Switching Operations}
Energy savings via cell-switching increase with the number of BSs switched off~\cite{bs_power,gb_blacksea}, emphasizing the need to open up the network to enable new opportunities for switching off. The availability of network capacity is crucial for cell-switching operations for several reasons~\cite{gb_blacksea}:
\begin{itemize} \item Dynamic power consumption is higher at macro BSs (MBSs) compared to small BSs (SBSs), requiring the deactivation of numerous SBSs to compensate.
\item High user density limits switching off opportunities; studies show in scenarios with increasing density, all cell-switching algorithms tend to maintain all BSs active~\cite{gb_blacksea}.
\item With constant user density, increased network capacity leads to more cell-switching opportunities, as additional radio resources are available to support offloaded users. 
\end{itemize}

\subsection{The Multi-Tier Cell-Switching Concept}

The multi-tier cell-switching concept, illustrated in Fig.~\ref{fig:concept}, features four tiers. The first tier, Tier-I, comprises traditional TN with terrestrial BSs, including MBS and SBS. Assuming a control-data separation architecture (CDSA), SBSs handle data traffic under the coverage of an MBS, primarily managing control signaling. In this setup, traffic from deactivated BSs is offloaded within Tier-I, to MBS, thus limiting cell-switching operations to the capacity and coverage of MBS.

Tier-II includes UAV-BSs and intelligent reflective surfaces~(IRS), mounted on UAVs~(referred to as UAV-IRS), along with robotic aerial BSs (RABS). UAV-BSs are primarily deployed for pop-up networking and disaster scenarios, whereas UAV-IRS serves as a network relaying component.
Additionally, fixed IRS components on tall buildings enhance relaying between UAV-BSs and terrestrial BSs, as shown in Fig.~\ref{fig:concept}.
RABS, equipped with anchoring mechanisms, attach to structures like lampposts to provide continuous service without hovering, conserving energy and adapting to traffic changes~\cite{RABS}.

Tier-III leverages the innovative technology of HAPS, classified as HIBS, which supports cellular networks with super-macro BS (SMBS)~\cite{haps_smbs}.
Positioned in the stratosphere at altitudes of 20 to 50 km, HAPS maintains a fixed position relative to the Earth, providing extensive coverage essential for 6G's ubiquitous connectivity.
Its capability for sustained operation, evidenced by maintaining a high-altitude balloon for over a year without landing or refueling, highlights its
potential for long-term applications like Earth observation,
and environmental monitoring. Although HIBS is gaining
recognition for cell-switching \cite{maryam}, ongoing development is required to fully realize its potential in this role.

Finally, Tier-IV encompasses satellite network, which consists of three orbits: geostationary Earth orbit (GEO), medium Earth orbit (MEO), and low Earth orbit (LEO).
With their extensive coverage footprints, satellites play a crucial role in NTN by providing connectivity where terrestrial and aerial alternatives are insufficient.
GEO satellites offer continuous service over fixed areas, making them suitable for stationary applications, while LEO satellites, with their lower development costs and reduced latency, enhance global connectivity through inter-satellite links (ISLs)~\cite{8643836}.
In the proposed framework, satellites serve as a complementary tier in the multi-tier cell-switching strategy by offloading non-latency-sensitive traffic and ensuring robust service continuity in remote or disaster-affected areas. Although the inherent latency of satellites poses a challenge, the integration of LEO satellites—characterized by lower latency compared to GEO—enhances their feasibility for dynamic cell-switching operations. To the best of our knowledge, this is the first time in the literature that satellite networks have been proposed for cell-switching, highlighting their potential to optimize energy efficiency while expanding coverage and capacity.

In the proposed multi-tier framework, cell-switching optimizes network efficiency by dynamically reallocating traffic across different tiers based on demand, energy efficiency, and latency requirements. Initially, traffic from deactivated SBSs is offloaded within Tier-I to the MBS. If further offloading is needed, higher tiers such as UAVs, HAPS, and satellites provide extended coverage and capacity. This tiered approach contributes to energy savings by turning off power-intensive BSs in lower tiers during low-traffic periods while utilizing energy-efficient NTN elements, such as HAPS powered by solar energy, to maintain service continuity. By making efficient use of available resources, the network minimizes unnecessary power consumption while maintaining an acceptable QoS.

Table~\ref{tab:Tiers} provides a comparative overview of each tier, highlighting key features such as capacity, coverage area, latency, sustainability, and deployment costs. Sustainability, in this context, encompasses three key dimensions: environmental, economic, and social. Environmentally, it refers to the reduction of energy consumption and carbon emissions through efficient resource utilization and the adoption of renewable energy sources where applicable. Economically, it involves optimizing infrastructure investments and operational expenditures to ensure long-term cost-effectiveness. Socially, sustainability focuses on improving network accessibility, bridging the digital divide, and supporting critical applications such as emergency response and rural connectivity.

\subsection{Latency as a Player: A New Character Unlocked}
The ITU's vision document for 6G introduces advancements beyond 5G, such as the evolution of ultra-reliable low-latency communication (URLLC) into hyper-reliable low-latency communication (HRLLC)~\cite{itu_vision_november}.
Thus, latency becomes crucial in 6G networks, necessitating its integration into cell-switching strategies not only for sustainability but also for HRLLC compliance.
While incorporating all tiers in cell-switching process seems beneficial, it introduces the challenge of increased latency, particularly in higher tiers like Tier-III and Tier-IV~\cite{ntn}, which must be managed to maintain QoS.
\begin{table}[t!]
\caption{Comparison of Different Tiers in Multi-Tier Cell-Switching}
\centering
\includegraphics[width=0.45\textwidth]{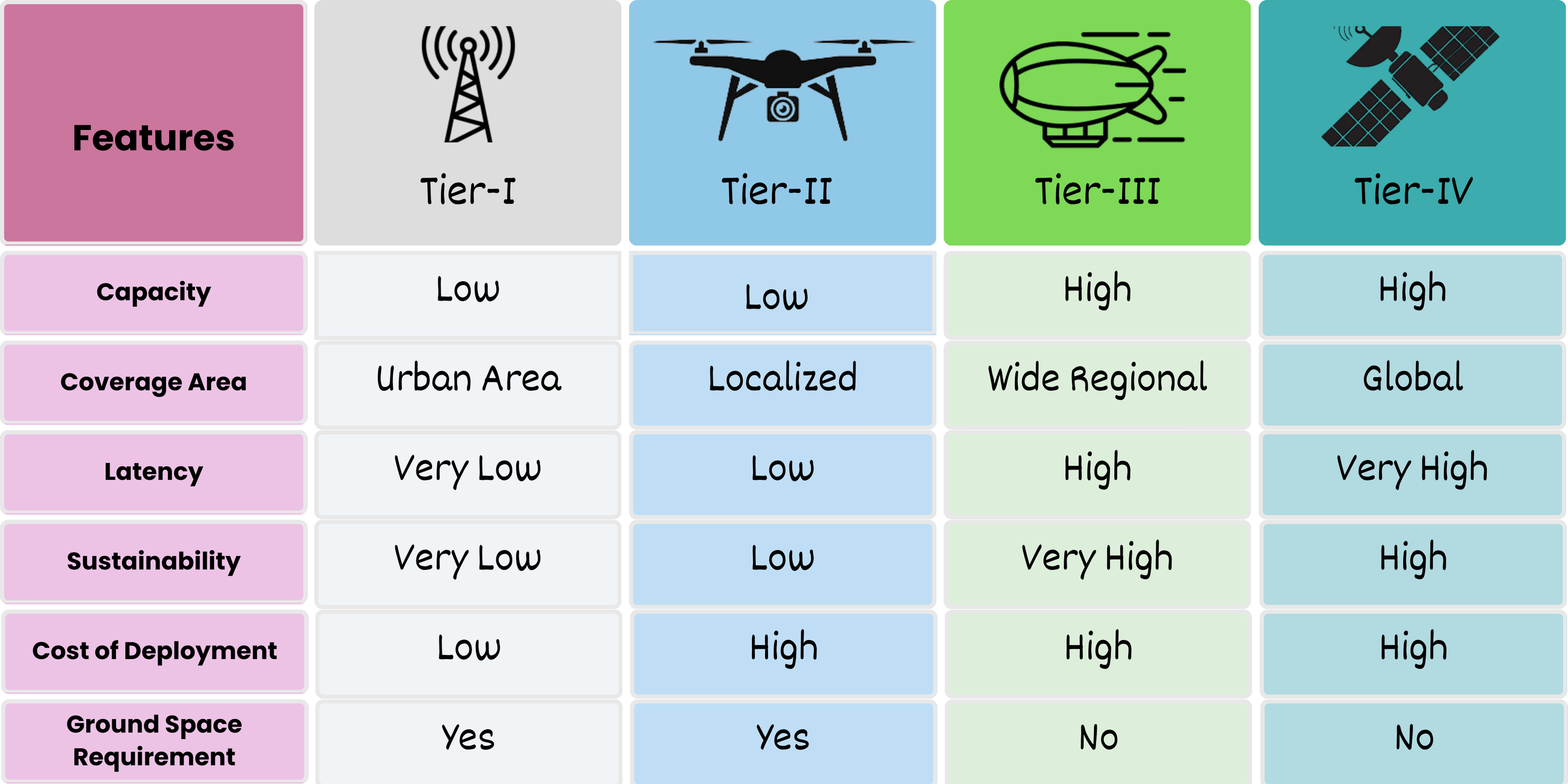}
\label{tab:Tiers}
\end{table}

To address latency concerns within the multi-tier cell-switching framework, employing context awareness is crucial. By categorizing users based on their latency profiles, the system enables informed cell-switching decisions. Highly delay-intolerant users, typical in HRLLC applications, are optimally managed within Tier-I, the tier with the lowest latency impact. Incorporating latency considerations into our approaches ensures adaptive measures when Tier-I reaches capacity, maintaining QoS. This adaptation is part of our approach, demonstrating that specific latency impacts can dictate targeted cell-switching strategies.

\subsection{Operational Procedures}\label{subsec:Procedure}

\begin{figure*}[t!]
    \centering
    \includegraphics[width=.65\textwidth]{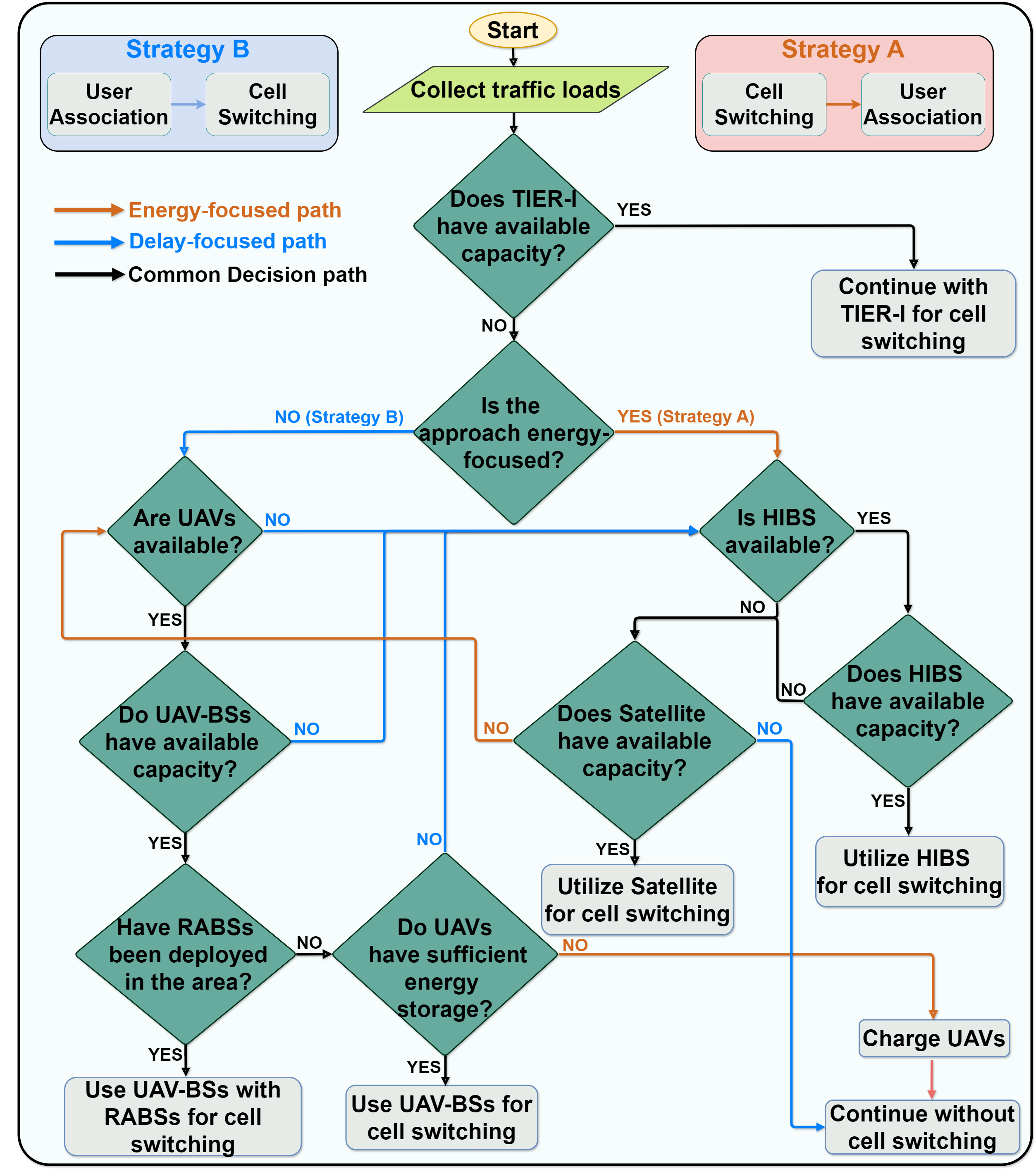}
    \caption{Conceptual frameworks for multi-tier cell-switching.}
    \label{fig:flowchart}

\end{figure*}

The proposed multi-tier cell-switching concept utilizes two tailored approaches, each addressing specific network requirements: an energy-focused approach prioritizing energy efficiency, and a delay-focused approach for minimizing latency in delay-sensitive applications.
More specifically, in the former (energy-focused approach), the additional energy consumption of the network is minimized, whereas in the latter (delay-focused approach), user satisfaction regarding latency requirements is prioritized.
In this regard, while the energy-focused approach favors offloading options/tiers that result in lower additional energy consumption, the delay-focused approach prioritizes network tiers that are closer to the ground, incurring lower latency.
Figure~\ref{fig:flowchart} integrates both approaches, incorporating two different paths (i.e. energy-focused and delay-focused paths) to illustrate the operation flow for each approach.
The energy-focused approach, detailed in Fig.~\ref{fig:flowchart}, begins with cell-switching to minimize the network's power consumption before associating users (i.e., Strategy A), which might occasionally compromise users QoS in terms of experienced delay.
The cell-switching process is driven by a centralized control system, likely located at a high-level operational center such as HAPS. This system employs real-time analytics and control channel mechanisms similar to the 5G New Radio (NR) Physical Downlink Control Channel (PDCCH) to ensure optimal resource allocation and efficient control signaling~\cite{8885990}. As illustrated in Fig.~\ref{fig:flowchart}, the algorithm initially checks for available capacity at the MBS (Tier-I), which is considered the most feasible option due to its proximity to users, offering better signal quality and requiring no additional deployment. 

If Tier-I is at full capacity, the preferred alternative is Tier-III over Tier-II, due to UAVs' limited battery life affecting their operational time and efficiency in communications.
Conversely, HIBS, positioned at higher altitudes, provides broader coverage and more reliable line-of-sight (LoS) connections than UAVs, and less delay compared to satellites.
Moreover, HAPS utilizes solar energy, boosting operational sustainability and aligning with our energy conservation goals. Typically active for various functions, not just cell-switching, HAPS eliminates the need for pop-up deployment, unlike UAVs. 
Thus, HAPS is prioritized in our energy-focused strategy for its sustained availability and multifunctional capabilities.
If capacity at HAPS is exhausted, satellites are then utilized, offering continual service without new deployments. If satellites reach full capacity, UAVs, with adequate energy and available RABSs are employed for cell-switching.

Conversely, the delay-focused approach, shown in Fig.~\ref{fig:flowchart}, is tailored to minimize latency for delay-sensitive applications. 
This approach emphasizes optimizing user association to enhance QoS and meet delay expectations, thus prioritizing user satisfaction over power conservation (Strategy B). 
In the context of this study, QoS is primarily defined by latency, which directly impacts user experience. The delay-focused approach ensures that applications requiring URLLC receive the necessary service levels by strategically offloading traffic to the most suitable tiers with minimal delay. Ensuring QoS involves maintaining low latency during cell-switching, optimizing user association policies, and minimizing disruptions across tier transitions.
Consequently, it may lead to suboptimal power consumption as the focus shifts from energy efficiency to ensuring user QoS.
The process starts with the same initial steps as the energy-focused approach. 
If Tier-I cannot accommodate additional traffic, offloading targets UAVs, employing those with sufficient energy and RABSs to minimize latency. 
If UAV capacity is unavailable, offloading proceeds to Tier-III and, if necessary, to Tier-IV to sustain cell-switching operations.

In short, the energy-focused approach prioritizes the network tiers that result in lower additional energy consumption, while the delay-focused approach prioritizes those closer to users to minimize round-trip delay.
The energy-focused approach reduces the additional energy consumption of the network by trying to utilize network components that are already available to eliminate static or deployment energy consumption.

\subsection{Multi-Tier Integration for Seamless Connectivity}

Integrating multiple network tiers presents a transformative opportunity to enhance network efficiency, enable seamless user experiences, and support direct-to-device communication in future networks. This integration can follow two potential paths: one where all layers operate within a unified ecosystem, and another where layers function across different ecosystems.
In the first scenario, all network layers including NTN, would share frequencies and standards under the 3rd Generation Partnership Project (3GPP) framework. This would allow users to remain unchanged, treating HAPS and satellites as extensions of terrestrial BSs with much larger coverage areas. Direct-to-device communication becomes feasible without needing additional transceivers, simplifying the user experience. LEO satellites could seamlessly support cell-switching operations without requiring devices to switch between different communication modes. The primary challenge in this harmonized setup is managing interference due to the vast coverage areas of NTN, which will necessitate advanced spectrum sharing and resource management techniques.

On the other hand, if NTN layers operate outside 3GPP standards, users would require modifications, evolving from the current dual-mode (cellular and Wi-Fi) setup to a three-mode system to support direct satellite/HAPS communication. This introduces inter-frequency handovers, increasing latency and complicating seamless communication when offloading traffic between TN and satellites. 


\section{Artificial Intelligence: A Savior or Guilty?}
 Cell-switching has predominantly been approached as a combinatorial optimization problem in the literature~\cite{ maryam,gb_blacksea}, targeting the optimal switching policy from a set of possible combinations. This approach faces two primary issues: 

\begin{itemize}
    \item Conventional optimization algorithms, like heuristics, assume prior knowledge of the network environment. However, wireless networks are highly unpredictable due to stochastic factors such as noise and shadowing, making these assumptions unrealistic. 

    \item These algorithms lack adaptive learning capabilities, making them unsuitable for dynamic conditions where environmental changes are frequent. 
   \end{itemize}

On the other hand, machine learning~(ML) or AI offers a practical solution for the cell-switching problem. 
Advancements in computational power and increase in data availability have significantly enhanced AI and ML. Modern processors, such as central processing units~(CPUs) and graphics processing units~(GPUs), have accelerated the implementation of ML algorithms, making them more efficient. The growth of digitalization and technologies like the Internet of things~(IoT) and social media platforms have also increased data accessibility. This convergence of advanced
computing capabilities and abundant data support the
practical and effective use of AI/ML in diverse applications,
including those in wireless networks.

\begin{figure}[t!]
    \centering
    \includegraphics[width=.47\textwidth]{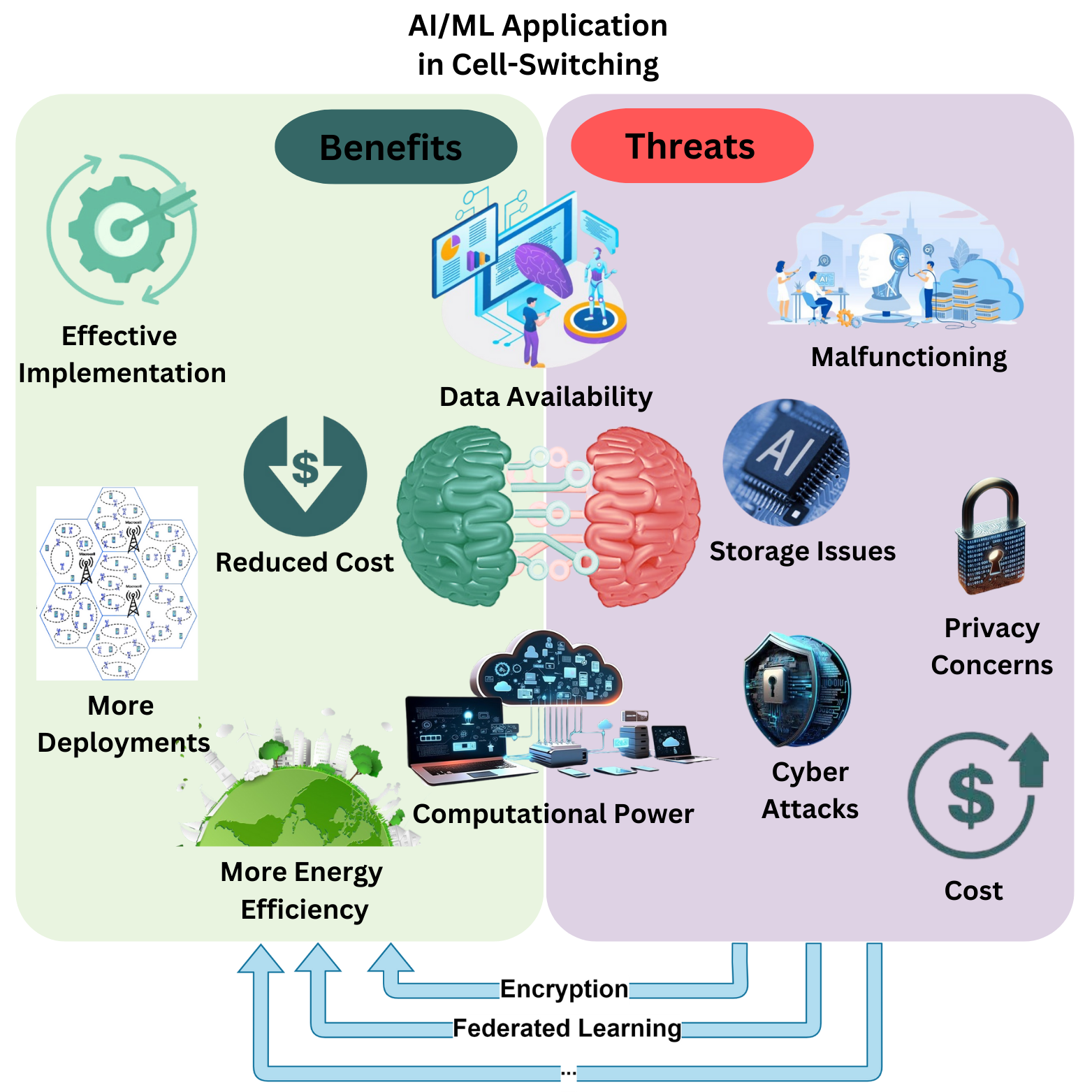}
    \caption{The benefits and the threats of AI/ML utilization in cell-switching problem.}
    \label{fig:ai_threats_benefits}
    \vspace{-0.5cm}
\end{figure}

While AI offers many benefits, security and privacy concerns are significant when deploying AI-based solutions~\cite{ai_security}. The abundance of data that enables AI's effectiveness also introduces risks of security breaches. Using ML algorithms requires careful consideration in protecting user-specific data such as profiles and locations. Preventive measures like cryptography, quantum computing, and federated learning are essential to safeguard data integrity~\cite{ai_security}. Figure~\ref{fig:ai_threats_benefits} shows both the benefits and the threats of AI/ML utilization in the cell-switching problem.

\subsection{Can Generative AI Play a Role?}
GenAI stands out in the AI landscape for its ability to create new content by learning complex patterns from large datasets. It goes beyond standard prediction, engaging in creative synthesis and innovation. GenAI converts various data types, from text to visuals, into efficient outputs, facilitating dynamic solutions across different fields.

In cell-switching applications, GenAI’s data compression and transformation capabilities optimize spectral efficiency and reduce network congestion. By converting bandwidth-intensive formats, such as video, into more manageable forms, GenAI minimizes transmission overhead while preserving essential information, reducing latency and improving overall network performance. This transformation lowers the number of resource blocks required per user and alleviates the traffic load on BSs~\cite{GenAI}. 
When the traffic loads of BSs are reduced in this way, energy saving can be increased by: (i) reducing dynamic load-dependent power consumption at BSs; (ii) increasing the likelihood of switching off the BSs as the traffic load and the probability of switching off are inversely proportional; and (iii) opening up more capacity to offload traffic from the switched off BSs.
Gen-AI frameworks can be implemented on BSs and user equipment using large-language models and on-device intelligence technology, respectively.

Beyond traffic offloading, variational autoencoders (VAEs) play a crucial role in network data transformation by learning compact representations of high-dimensional traffic patterns. This enables bandwidth-efficient transmission and enhanced anomaly detection, facilitating adaptive resource management for energy-efficient switching across tiers.
Additionally, transformer-based models further optimize handover efficiency by predicting user mobility trends and dynamically adjusting traffic transformation strategies based on real-time network conditions. These models refine resource allocation by identifying congestion-prone areas and balancing traffic flow, ensuring seamless transitions between TN and NTN layers.
Device compatibility is another domain where GenAI proves beneficial. By enabling virtualization through software-defined networking (SDN), GenAI allows devices to seamlessly switch between TN, HAPS, and satellite communications without physical modifications, enhancing adaptability across different network protocols.
In our framework, VAEs improve spectral efficiency by encoding network data into compact representations, reducing redundant information while enhancing energy savings. Additionally, transformer-based models optimize handovers by predicting mobility trends and dynamically adjusting traffic transformation strategies, ensuring seamless TN-NTN transitions.

\section{Case Study: Cell-switching with the Availability of Multi-Tiers}\label{sec:case_study}

To evaluate the effectiveness of our proposed multi-tier cell-switching strategy, we conduct a comprehensive simulation study focusing on two key performance metrics: network power consumption and user dissatisfaction. The first metric assesses energy efficiency of different tier configurations, while the second quantifies the impact of cell-switching on user experience, particularly for latency-sensitive applications. The trade-off between these metrics is evident: the energy-focused approach maximizes energy savings, whereas the delay-focused approach prioritizes minimizing user dissatisfaction.
The evaluation compares two network operation strategies: the conventional all-active approach (A3) and proposed exhaustive search-based approach (ES). In the A3 approach, all BSs remain active throughout simulation, leading to high power consumption but ensuring full-service availability. In contrast, the ES approach systematically explores all possible ON/OFF configurations for the SBSs to determine optimal state vector that minimizes network power consumption. While ES provides optimal solution for cell-switching, the novelty of our work lies in multi-tier cell-switching framework rather than the choice of a specific cell-switching method.

To analyze these strategies, we model a four-tier vertical heterogeneous network deployed in a $1 \text{km}\times 1\text{km}$ urban area. The TN consists of Tier-I, which includes six SBSs—two RRHs, two microcells, one picocell, and one femtocell—each with distinct power profiles, alongside an MBS, as detailed in~\cite{9344664}. These SBSs are symmetrically deployed within the coverage area to ensure balanced network distribution. The aerial network comprises Tier-II, incorporating a UAV-BS, and Tier-III, which features a HAPS. The space network (Tier-IV) integrates a LEO satellite, which provides additional offloading capacity.
Simulation procedure follows a structured approach, incorporating user mobility, traffic demand, and dynamic association mechanisms. Users are uniformly deployed in urban area and initially associated with the nearest SBS based on a free-space path loss model. Users are categorized based on mobility type (pedestrian, cyclist, and vehicular) and delay tolerance (intolerant, mid-tolerant, and tolerant). During each simulation time slot, users generate traffic demands following a uniform discrete distribution within a predefined range. Once the initial association is established, cell-switching is performed based on ES calculations, and traffic offloading is executed according to the operational procedure described in Section~\ref{subsec:Procedure}, depending on the applied approach.
The power consumption of each SBS is initially calculated using the EARTH model~\cite{6056691}, and multi-tier cell-switching strategies are applied to assess different tier configurations. 

\begin{figure}[t!]
    \centering
    \includegraphics[width=.45\textwidth]{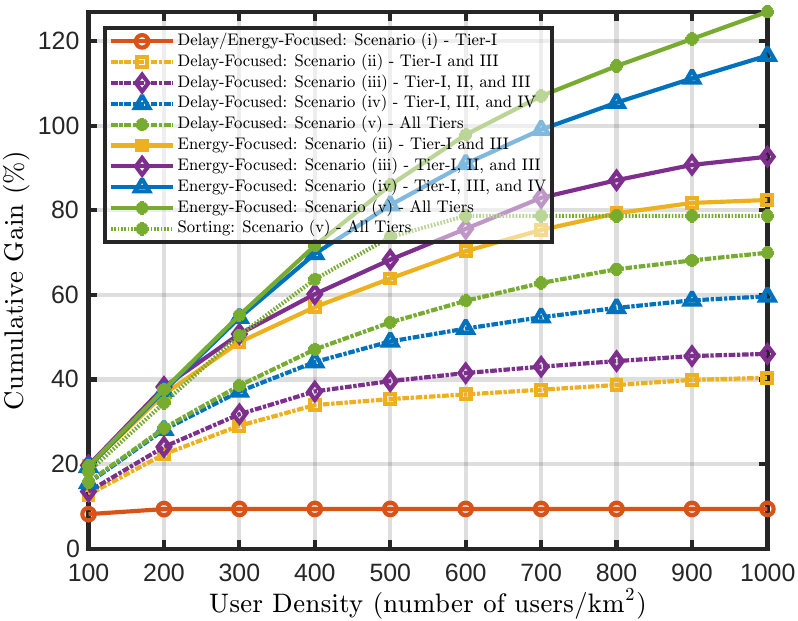}
    \caption{Power consumption gains across various tiers at different user densities.}
    \label{fig:Gain}
\end{figure}

Figure~\ref{fig:Gain} shows cumulative power consumption gains across user densities for both energy-focused and delay-focused approaches. These gains are compared with conventional A3, assuming a 50\% availability of HIBS for cell-switching. The cumulative gain metric quantifies energy efficiency improvement by measuring the reduction in total power consumption achieved through ES cell-switching, relative to A3 baseline.
We evaluate five scenarios: (i) Tier-I; (ii) Tier-I and III; (iii) Tier-I, II, and III; (iv) Tier-I, III, and IV; and (v) All-Tiers. Additionally, we include the sorting-based greedy approach, where the SBSs are sorted based on their traffic loads and switched off until the capacity of offloading destination is full, for scenario (v) to demonstrate the effectiveness of ES in achieving optimal energy savings.
Results in Fig.~\ref{fig:Gain} show increasing power savings across user densities. At lower densities, multi-tier integration yields significant gains. However, as densities increase, the rate of gain increment diminishes. This slowdown is due to SBSs reaching full capacity, which limits effective cell-switching opportunities.
Moreover, the graph indicates minimal and identical gains from scenario (i) alone for both approaches since Tier-I is the sole offloading destination.
Scenarios (ii) and (iii) achieve better gains than scenario (i) due to the availability of aerial network tiers, including UAV and HAPS. However, as shown in the figure, the introduction of the satellite layer in scenario (iv) leads to a significant improvement in power savings, particularly at higher user densities. 
Despite their high altitude and long-distance links, satellites play a crucial role in power savings due to their extensive coverage and persistent availability for cell-switching. Compared to aerial tiers, satellites offer a large-scale offloading option, making them particularly effective in high-demand scenarios where terrestrial and aerial tiers reach capacity. The significant power savings observed in scenario (iv) compared to scenarios (i), (ii), and (iii) underscore the increasing benefits of multi-tier integration and the satellite’s critical role in extending network capacity and optimizing energy efficiency. As illustrated in the figure, the energy-focused approach achieves significantly higher gains by prioritizing network power reduction through extensive user offloading across multiple tiers.

\begin{figure}[t!]
    \centering
    \includegraphics[width=.45\textwidth]{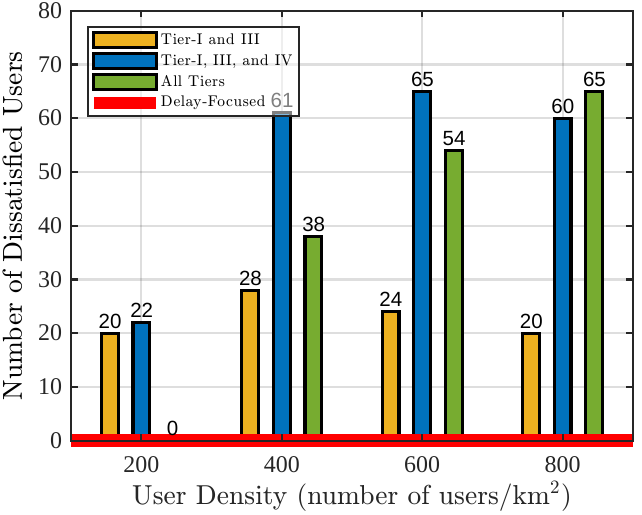}
    \caption{Distribution of dissatisfied users across different tiers at varying user densities for the energy-focused approach.}
    \label{fig:dissatisfaction}
\end{figure}

Although the energy-focused approach minimizes power consumption, it may compromise user experience, particularly in terms of delay. This trade-off becomes evident when cell-switching prioritizes energy efficiency over delay requirements. User dissatisfaction is measured as the number of delay-intolerant users offloaded to Tier-III or Tier-IV and mid-tolerant users offloaded to Tier-IV, as these offloading decisions exceed their acceptable delay limits.
Figure~\ref{fig:dissatisfaction} shows user dissatisfaction under the energy-focused approach, highlighting the delay-focused approach's effectiveness in minimizing latency and eliminating dissatisfaction for delay-sensitive applications.
In the energy-focused approach, when only Tier-I is used for offloading, all users, including the delay-intolerant ones, generally experience no dissatisfaction. This scenario is not shown in the figure.
In more complex multi-tier scenarios, dissatisfaction patterns vary. The second scenario (Tier-I, III, and IV) exhibits higher dissatisfaction than the first scenario (Tier-I and III) due to increased delays from offloading to satellites in Tier-IV. Conversely, the third scenario (All-Tier Integration) reduces dissatisfaction compared to the second, due to the lower delays of Tier-II.
Across all scenarios, dissatisfaction initially rises with user density due to more delay-intolerant and mid-tolerant users. However, as user density peaks, dissatisfaction declines as high-load SBSs limit cell-switching, reducing power consumption by keeping more SBSs active.

The combination of the energy- and delay-focused approaches can also be considered as an intermediate and alternative solution.
To achieve this, first, a softer approach is needed, such that the definition of satisfaction needs to be converted from its binary form (requiring a hard decision) to a continuous form, by considering the difference between the provided and requested delay values.
Then, partial offloading of users to different network components/tiers is allowed in the energy-focused approach to create a soft-type process rather than the presented hard-type process that requires architectural switches.
On the other hand, this requires further research, as the problem becomes more complex due to the increased number of constraints and variables that need to be optimized.

\section{Opportunities and Challenges}\label{sec:Opportunitiyes and Challenges}
The multi-tier cell-switching approach significantly enhances
network energy efficiency, coverage, and capacity,
offering a promising path toward sustainable and adaptive communication networks. However, realizing its full
potential requires overcoming several challenges. First, interference management remains a critical challenge due to coverage overlaps in multi-tier networks. This challenge is particularly pronounced in NTN, where extensive coverage and strong LoS links can lead to significant interference. Efficient radio resource management strategies, particularly for co-channel deployments, are essential to mitigate interference and ensure reliable service quality.
Second, adopting accurate channel modeling that accounts for shadowing and fading is essential, moving beyond oversimplified free-space path loss models. Integrating more realistic channel models will improve the accuracy of network planning and performance evaluation.
Third, the integration of GenAI introduces both opportunities and challenges. While GenAI can optimize cell-switching through data-driven decision-making, its implementation raises concerns regarding data privacy and security, as well as the computational cost of running complex AI models in real-time. Developing lightweight and privacy-preserving AI frameworks is essential for practical deployment.
Fourth, in addition to the individual and specialized frameworks mentioned in Section~\ref{sec:case_study}, building a framework that can be used to optimize energy and latency simultaneously is a challenging but necessary aspect.
Lastly, as IoT continues to expand across various sectors and becomes critical to seamless 6G connectivity, incorporating IoT device power consumption, particularly during uplink transmissions, must be considered to provide a comprehensive sustainability analysis of network operations.


\section{Conclusion}\label{sec:conclusion}
This article presents a comprehensive study of the multi-tier cell-switching strategy for sustainable wireless networks, significantly improving energy efficiency and network performance. By implementing cell-switching across network tiers with distinct capabilities, we achieved substantial reductions in energy consumption, enhancing efficiency and QoS. Two distinct approaches were introduced: an energy-focused approach optimizing power for sustainability, and a delay-focused approach minimizing latency for delay-sensitive applications, thereby improving QoS. The integration of AI/GenAI further improves adaptability and network efficiency, supporting more dynamic and optimized operations. 

\section*{Acknowledgment}

This study is supported in part by The Scientific and Technological Research Council of Turkiye (TUBITAK), and in part by the Connected and Autonomous Vehicles (TrustCAV) CREATE Program funded by the Natural Sciences and Engineering Research Council of Canada (NSERC).

\bibliographystyle{IEEEtran}
\bibliography{refs}

\section*{Biography}

\textbf{Metin Ozturk} (metin.ozturk@aybu.edu.tr) is currently an Assistant Professor at Ankara Yıldırım Beyazıt University, Turkiye. His current research includes intelligent networking for wireless communications, with a particular focus on non-terrestrial
networks (NTN) and sustainability.

\textbf{Maryam Salamatmoghadasi} (maryamsalamatmoghad@cmail.carleton.ca) is a
Ph.D. Candidate in the Non-Terrestrial Networks (NTN) Lab at Carleton
University, Canada. Her research focuses on the integration of High
Altitude Platform Stations (HAPS) with terrestrial wireless networks and sustainability.

\textbf{Halim Yanikomeroglu} (halim@sce.carleton.ca) is a Chancellor’s Professor in
the Department of Systems and Computer Engineering at Carleton University,
Ottawa, Canada, leading the Non-Terrestrial Networks (NTN) Lab. His
research group focuses on wireless access architecture for the 2030s and
2040s, with a particular emphasis on NTN.


\end{document}